\documentclass[11pt,twoside]{article}

\usepackage{asp2004}
\usepackage{epsf}
\usepackage{psfig}
\usepackage{lscape}

\begin{document}
\title{The Halo of the Milky Way}   
\author{Heidi Jo Newberg \& Brian Yanny}   
\affil{Department of Physics, Applied Physics and Astronomy, 
Rensselaer Polytechnic Institute, Troy, NY 12180; 
Fermi National Accelerator Laboratory, P.O. Box 500, 
Batavia, IL 60510}    

\begin{abstract} We show that the star counts in the spheroid of the Milky Way are 
not symmetric about the $l=0^\circ$, $l=180^\circ$ plane.  The minimum counts are found 
towards $l=155^\circ$.  The Galactic longitude of maximum star counts depends on the 
magnitude and color selection of the halo stars.  We interpret this as evidence that the
spheroid population is triaxial with a major axis oriented $65^\circ$ from the line of sight
from the Sun to the Galactic center, and approximately perpendicular to the Galactic bar.
Large local star concentrations from tidal debris and possible tidal debris are also observed.
A full understanding of the Galactic spheroid population awaits position information and
three dimensional space velocities for a representative set of stars in every substructure.
Tangential velocities for many stars will be provided by current and planned astrometry
missions, but no planned mission will measure stars faint enough to unravel the more 
distant parts of the spheroid, which contain the majority of the spatial substructure.
This paper uses data from the Sloan Digital Sky Survey (SDSS) public data release DR3.
\end{abstract}

\section{Introduction}
While the Milky Way galaxy gives us a much better view of the individual 
component stars than any other galaxy, it is difficult to view our galaxy as
an assembled whole.  New surveys such as 2MASS and SDSS \citep{yetal00,getal98}, 
 are allowing us to start assembling the global picture 
one star at a time.  Other surveys that are in progress such as RAVE, GAIA, 
Pan-STARRS, UKIDSS/VISTA, and SDSS II/SEGUE will extend our view of the 
stars not only in positional information but also in kinematic information 
for a large number of Galactic stars. 

\section{Evidence that the spheroid is not symmetrical}

SDSS color-magnitude Hess diagrams such as Figure 11 of \cite{netal02} show 
two populations with distinct turnoffs (see Fig. 1).  The brighter 
population has a turnoff at $g-r \sim 0.4$ and is associated with the 
thick disk.  The fainter population has a turnoff at $g-r \sim 0.3$, and 
is associated with the spheroid. Fig. 1 shows the color-magnitude selection 
boxes for thick disk, spheroid, and faint spheroid data samples
selected from SDSS-DR3.

Fig. 2 shows the number counts of spheroid stars as a function of position in
the Galaxy from SDSS DR3.  Where there are no counts, we have no data.  Note
that the number counts do not appear to be symmetric about the Galactic 
center.  Fig. 3 shows this numerically, by plotting star counts along several
lines of constant Galactic latitude.  There are more spheroid stars in
quadrant IV than in quadrant I, and the minimum is at $l=155^\circ$ rather than
$l=180^\circ$, as we would have expected.

Fig. 4 shows that the spheroid number counts are not significantly asymmetric about the
Galactic equator.  Fig. 5 shows the star counts for the thick disk selection
for all of DR3, while Fig. 6 shows quantitatively that the thick disk is
symmetric about $l=0^\circ,180^\circ$.

\section{Interpretation that the spheroid is triaxial}

Fig. 7 is a cartoon, looking down on the Galactic plane, depicting the star
counts that would be observed at a constant height above or below the 
Galactic plane for a sample triaxial spheroid.  This type of model can 
explain a minimum in quadrant II, and a maximum in quadrant IV for samples of
stars close to the Sun.  It predicts that the maximum would switch to
quadrant I for stars more distant than the distance from the Sun to the
Galactic center.

Fig. 8 shows the star counts for a more distant sample of spheroid stars, and
Fig. 9 shows star counts through this figure at constant Galactic latitude.
There is some evidence that the peak star counts shifts to quadrant I for 
this more distant sample, though the presence of the Sagittarius stream
makes the interpretation problematic.

\begin{figure}
\plotfiddle{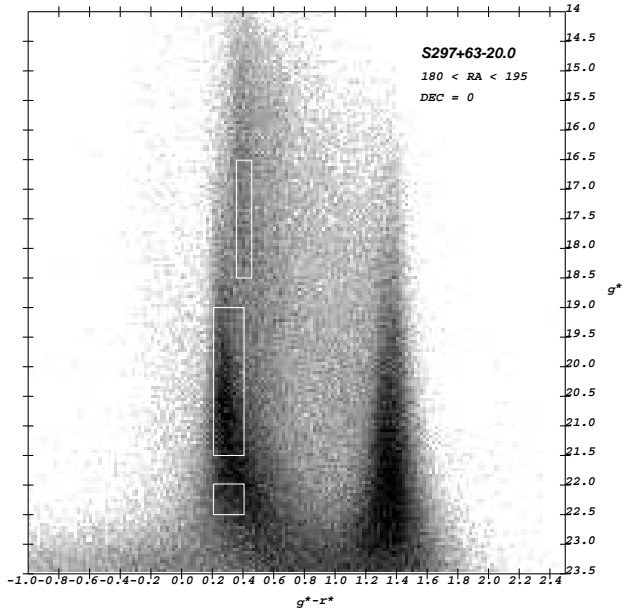}{2.85in}{0}{40.0}{40.0}{-150in}{-50in}
\caption{SDSS color-magnitude Hess diagram, reprinted from Figure 11 of \cite{netal02}.  
All magnitudes are reddening-corrected.
Overlaid are color-magnitude selection boxes for the thick disk (top), spheroid (middle)
and faint spheroid (bottom).  There was also a $u-g > 0.4$ color cut to exclude UV excess
quasars.}
\end{figure}

\begin{figure}
\plotfiddle{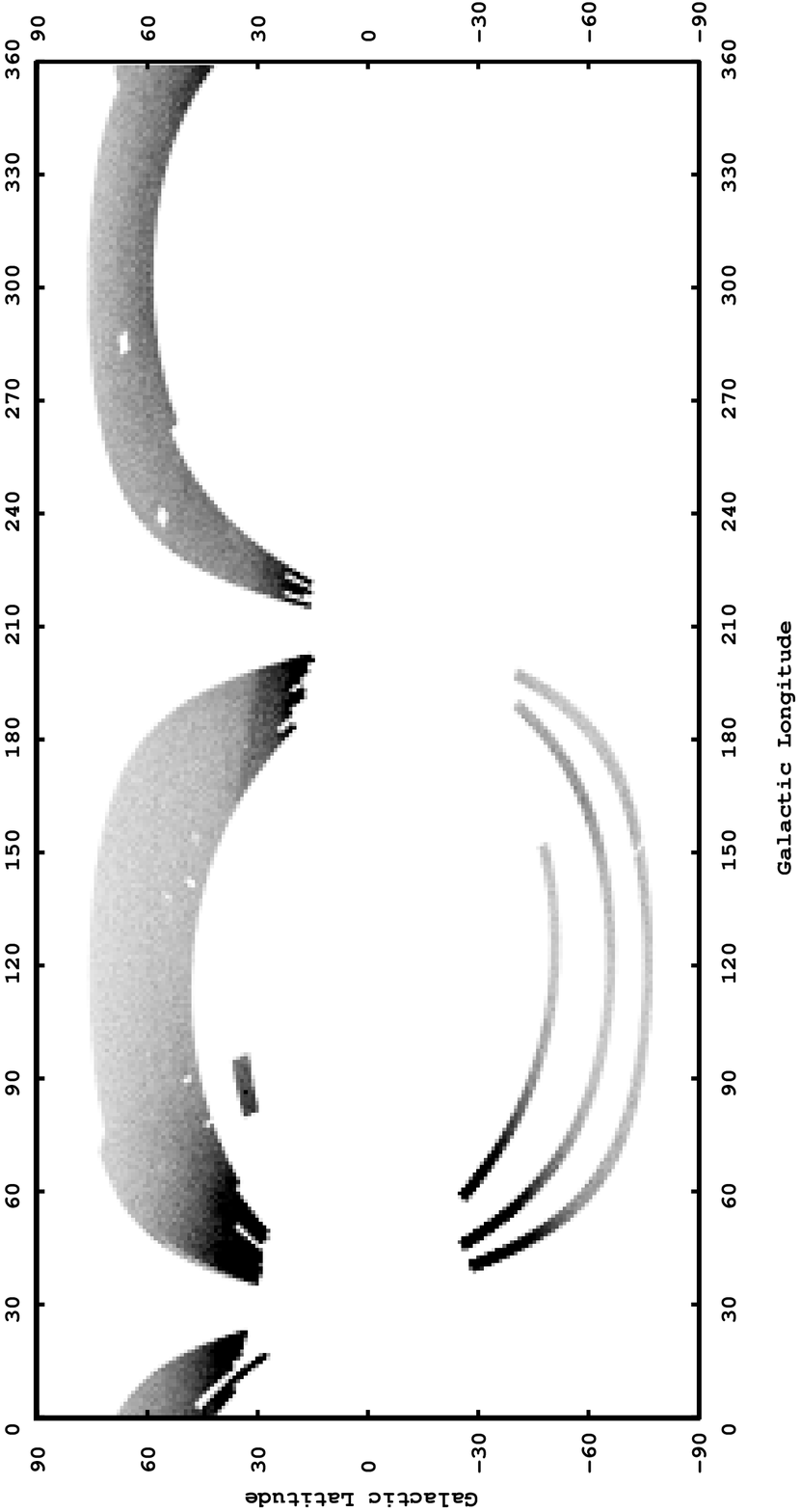}{7.25in}{0}{72.0}{72.0}{-250in}{-40in}
\caption{
Density of spheroid stars in the sky from SDSS DR3, shown in Galactic coordinates.  The densities have not been corrected for the effects of $\cos(b)$.  Notice that the density
appears smooth - no obvious discontinuities at stripe/run boundaries.  Also, the distribution
does not appear be be symmetric about $l=180^\circ$.
}
\end{figure}

\begin{figure}
\plotone{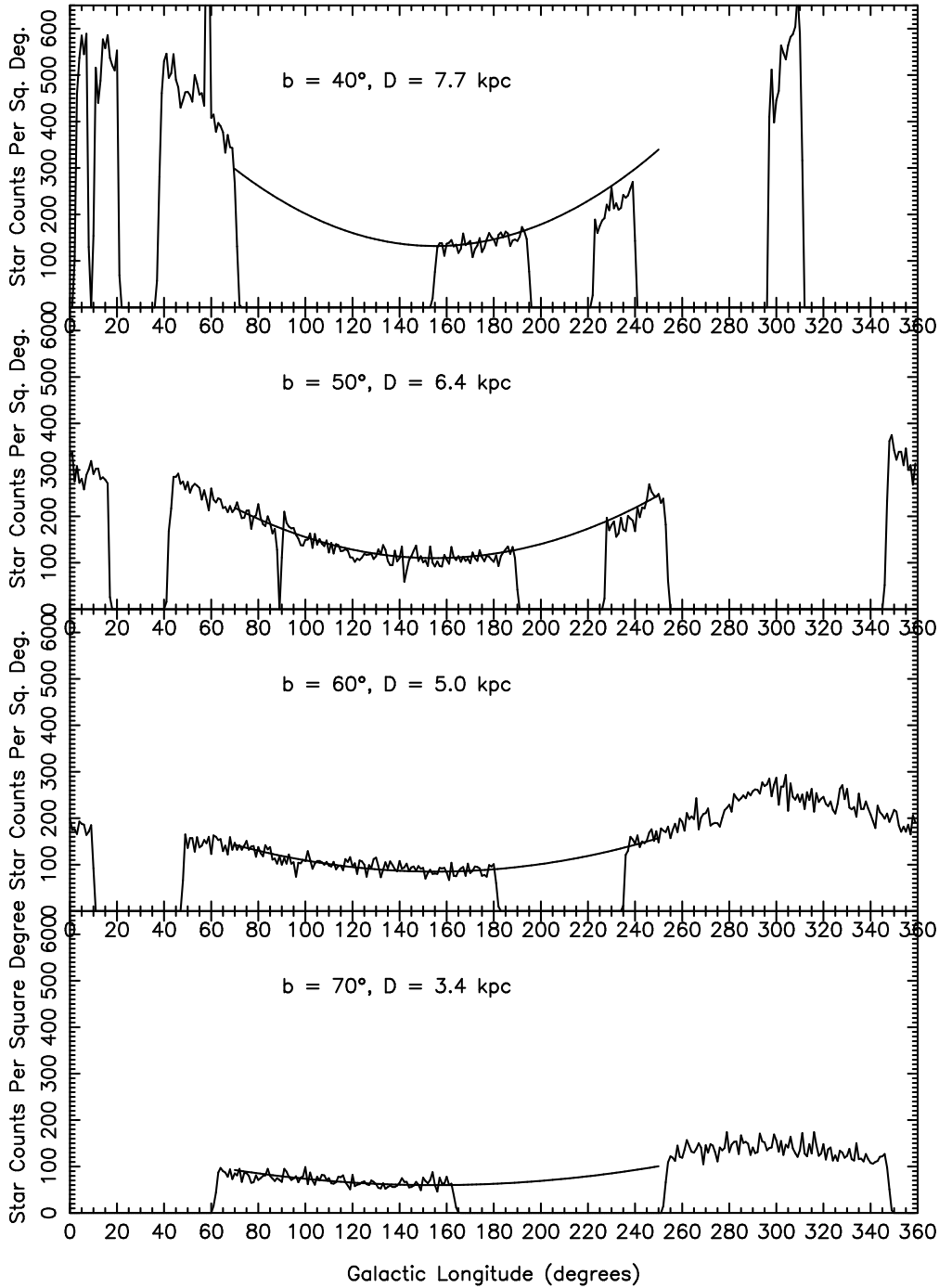}
\caption{Star counts at constant Galactic latitude.  These are plots of the data in Figure 2, 
along four Galactic latitudes, uncorrected for $cos(b)$ so the
``star counts per sq. deg." labels are a bit misleading.  Parabolas with minima at
$l=155^\circ$ have been drawn over the data.  Spheroid models that are not triaxial would
predict a minimum at $l=180^\circ$.  There is also a maximum in the star counts 
near $l=300^\circ$, not at $l=0^\circ=360^\circ$.}
\end{figure}

\begin{figure}
\plotone{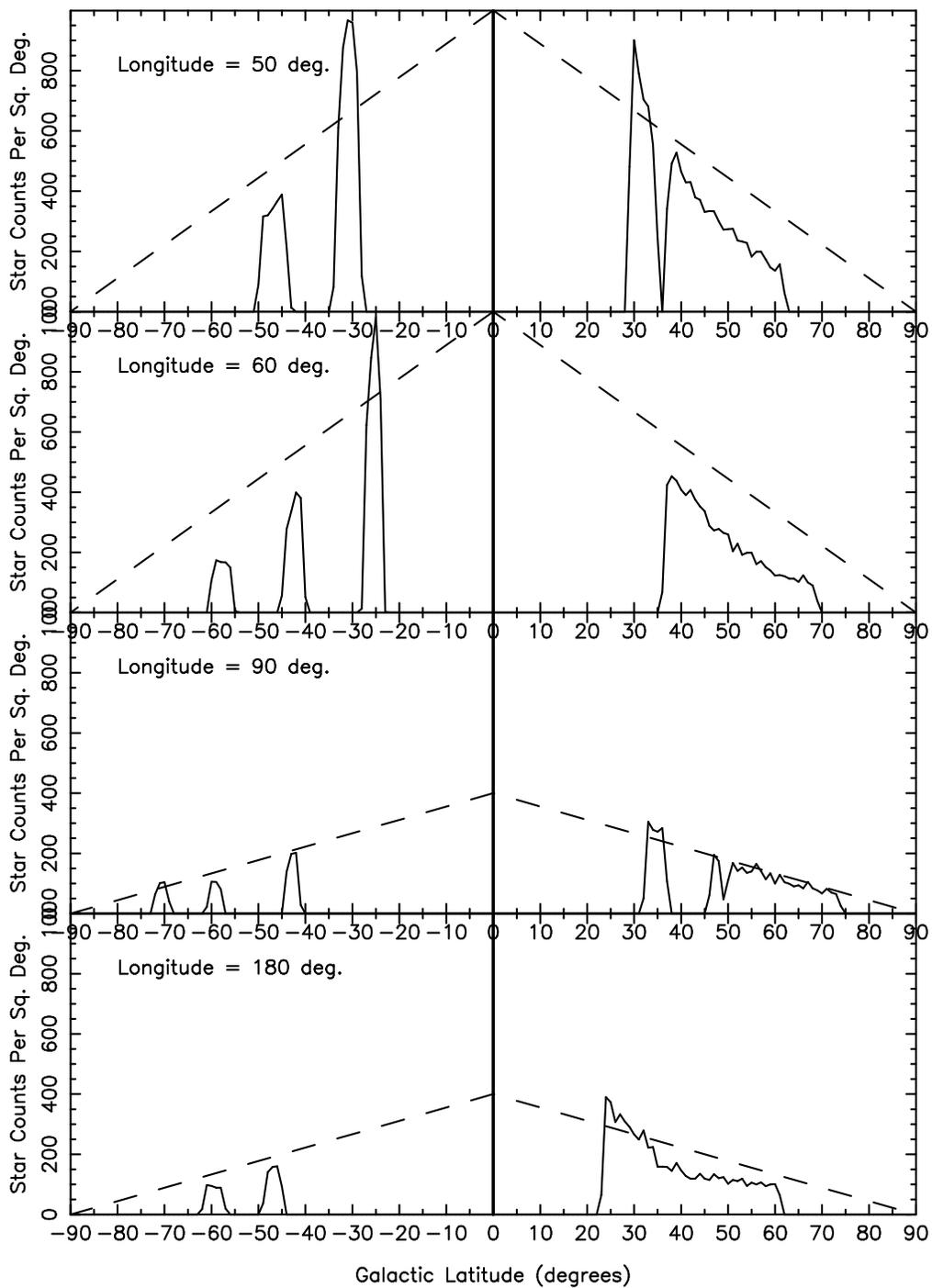}
\caption{Star counts at constant Galactic longitude.  These are plots of the data in Figure 2, along four Galactic longitudes.  The dotted lines are drawn to help the eye ascertain the symmetry of the data about the Galactic equator.}
\end{figure}

\begin{figure}
\plotfiddle{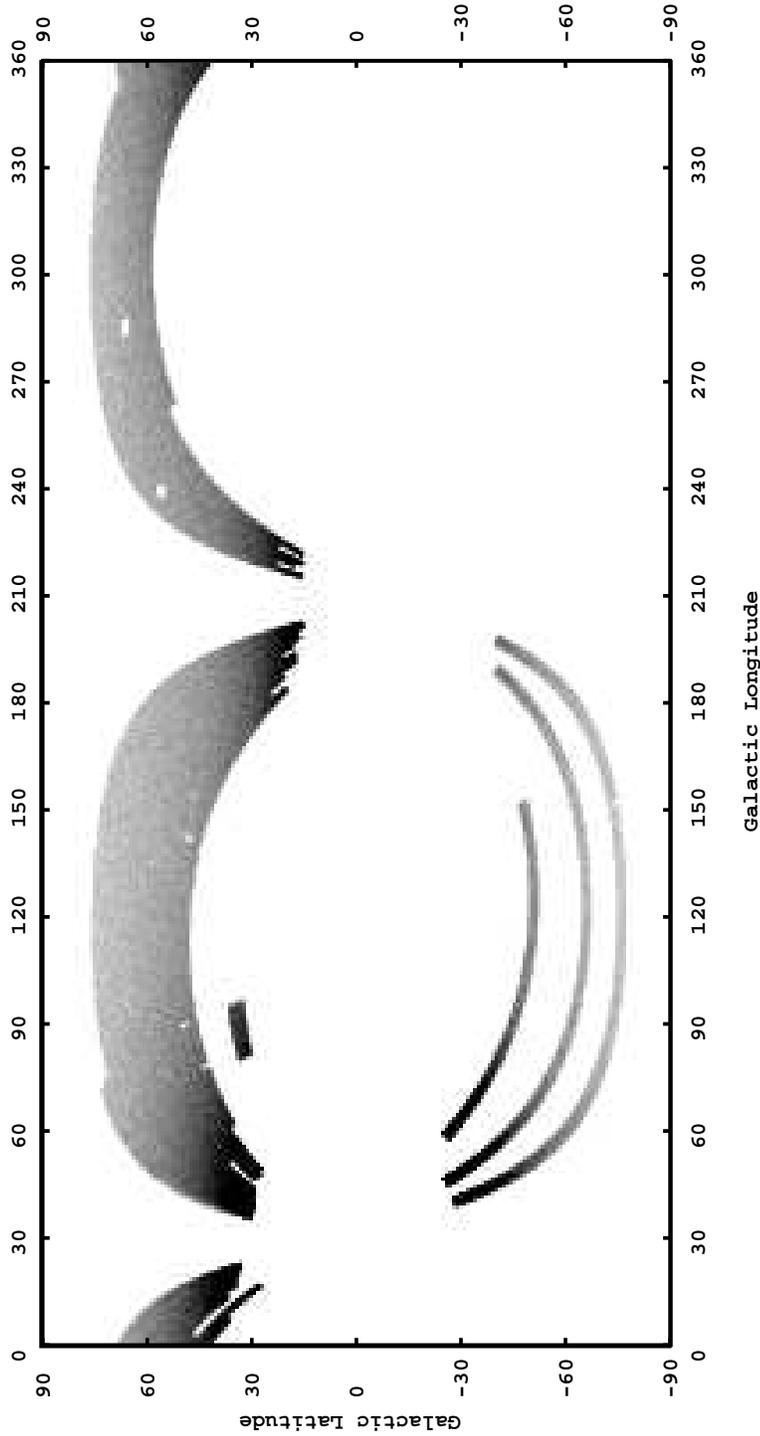}{7.25in}{0}{72.0}{72.0}{-250in}{-40in}
\caption{
Density of thick disk stars in the sky from SDSS DR3 (see selection box from Fig. 1), shown in Galactic coordinates.  The densities have not been corrected for the effects of $\cos(b)$.  Notice that the density
appears smooth - no obvious discontinuities at stripe/run boundaries.  Also, the distribution
appears be be symmetric about $l=180^\circ$.
}
\end{figure}

\begin{figure}
\plotone{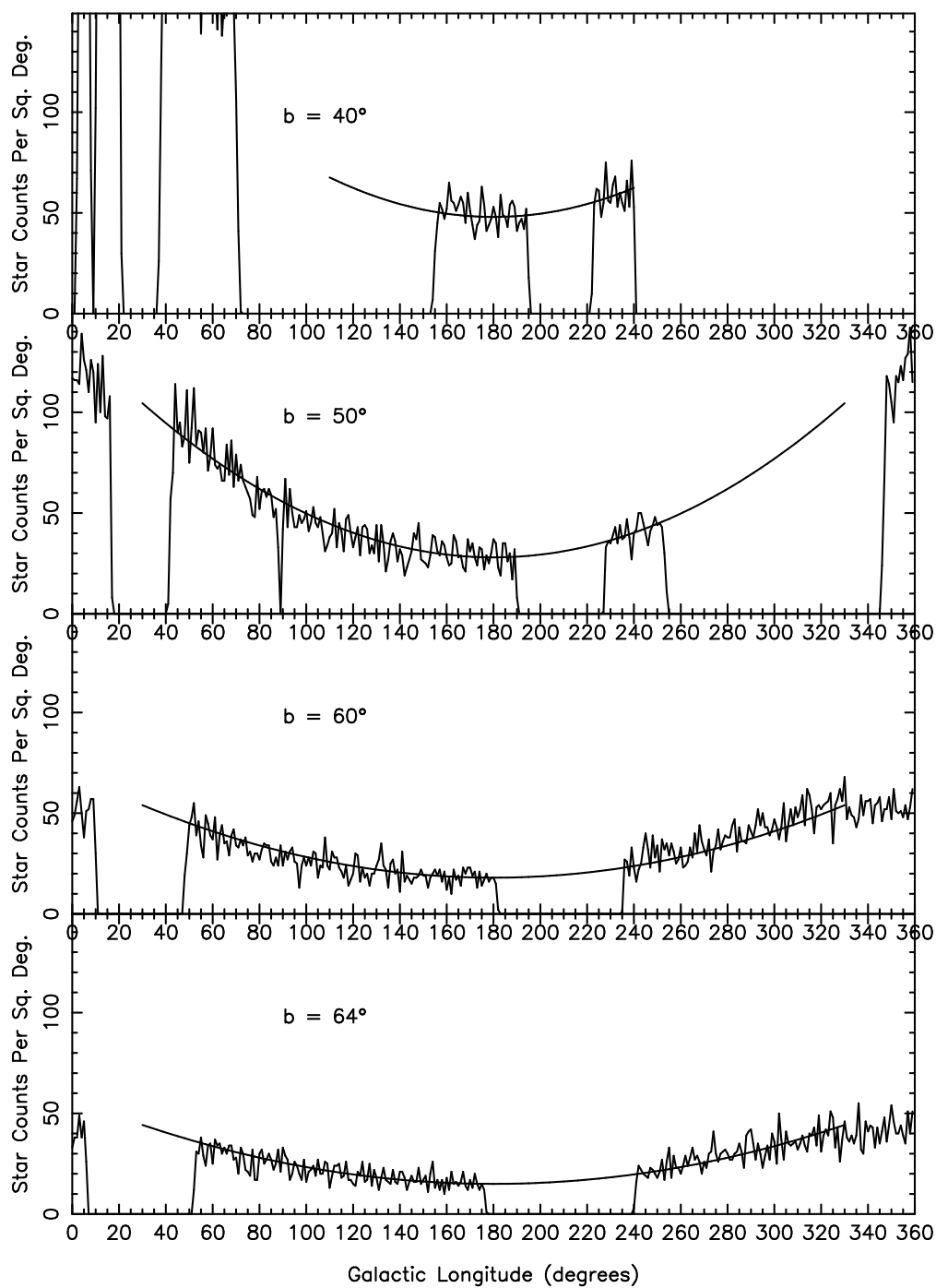}
\caption{
Thick disk star counts at constant Galactic latitude.  These are plots of 
the data in Figure 5, along four Galactic latitudes.  These also have not 
been corrected for $cos(b)$, so the labels as star counts per square degree 
is a bit misleading.  Parabolas with minima at $l=180^\circ$ have been 
drawn over the data.  The maximum star density is at $l=0^\circ$, as we
expect.}
\end{figure}

\begin{figure}
\plotfiddle{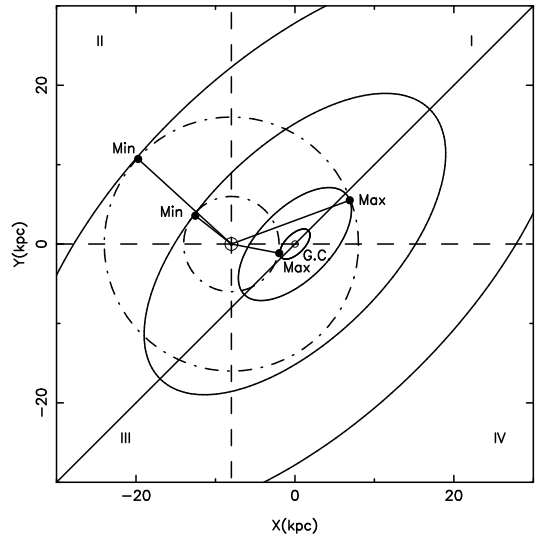}{2.5in}{0}{40.0}{40.0}{-150in}{-50in}
\caption{
Iso-density star counts for a triaxial spheroid as observed from the Sun.  
Solid lines show isodensity star counts for a triaxial spheroid with major
axis $45^\circ$ degrees from our line of sight, projected on the Galactic
plane.  Dashed lines show two projected distances from the Sun. 
For close stars, the maximum density is in quadrant
IV and the minimum is in quadrant II.  For distances larger than the
Sun-GC distance, the maximum is in quadrant I and the minimum is in quadrant
II.  For large distances, the minimum is perpendicular to the major axis 
of the spheroid.}
\end{figure}

\section{Conclusion}

The spheroid population may be triaxial with a major axis oriented 
$65^\circ$ from the line of sight from the Sun to the Galactic center.
This is approximately perpendicular to the Galactic bar.  This is surprising
since one typically expects bars to have an angular pattern speed larger than
could be sustained in the spheroid, but the possible connection should be
examined. 

Significant substructure is observed in the spheroid population with main
sequence turnoffs at apparent magnitudes of $g=20$ and fainter.  Even
the GAIA mission will not reach faint enough or distant enough
to study the part of the spheroid
where we see the most substructure.  Thus, the need for newer and better
astrometry will be required for the indefinite future.

\begin{figure}
\plotfiddle{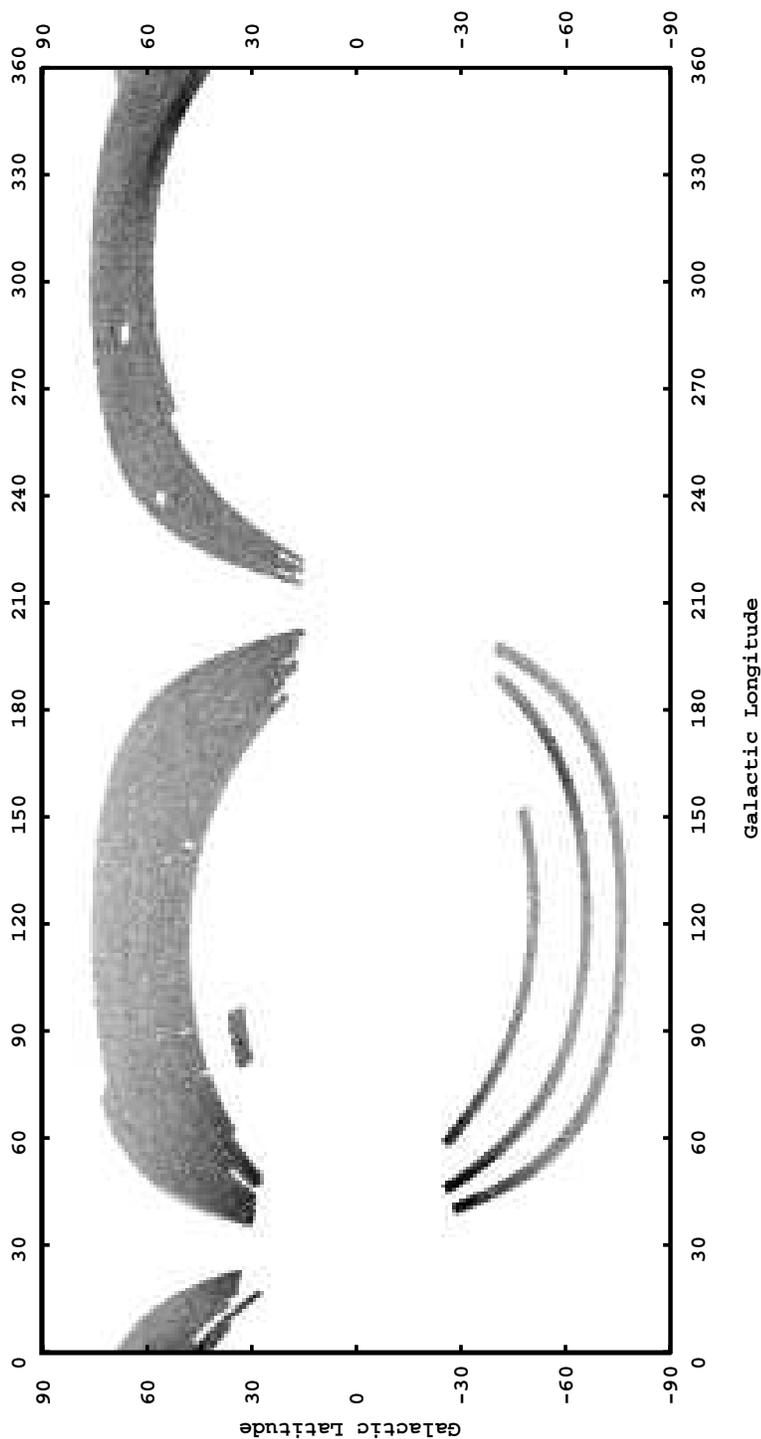}{7.25in}{0}{72.0}{72.0}{-250in}{-40in}
\caption{
Density of faint spheroid stars in the sky from SDSS DR3 (see selection box 
from Fig. 1), shown in Galactic coordinates.  The densities have not been 
corrected for the effects of $\cos(b)$. 
The distribution is not symmetric about $l=180^\circ$.  The Sagittarius
tidal stream is clearly visible within $40^\circ$ of $l=340^\circ$.}
\end{figure}

\begin{figure}
\plotone{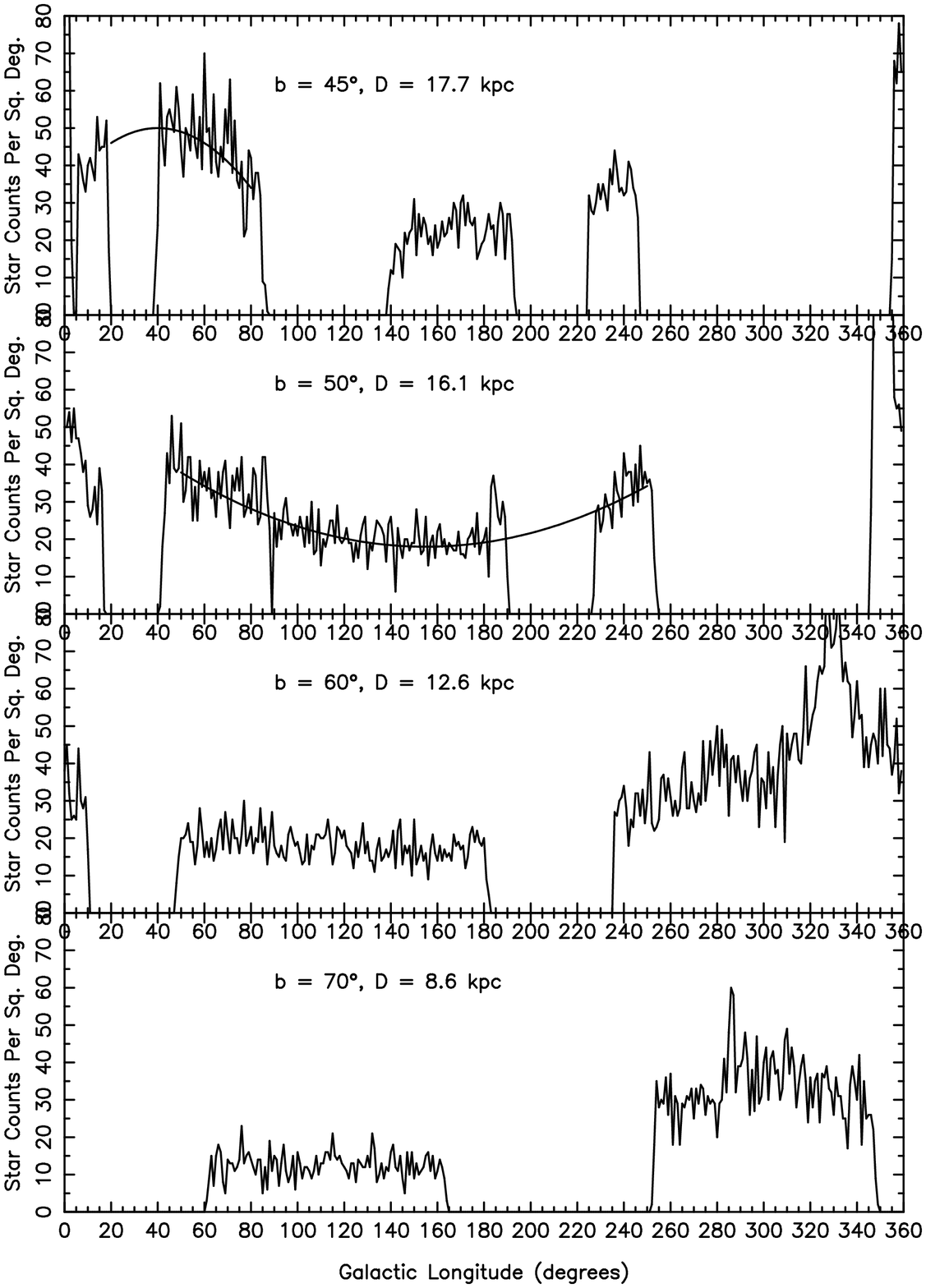}
\caption{Faint spheroid star counts at constant Galactic latitude.  These 
are plots through the data in Fig. 8.
A parabola with minimum at $l=155^\circ$ has been drawn over the second panel. 
The star counts on the right half
of the plots show the stars from the Sgr dwarf tidal stream.
The top panel, which gives the densities of spheroid stars with the largest
distances from the Sun as projected on the Galactic plane, shows a peak in 
the star counts at $l=40^\circ$.}
\end{figure}

\acknowledgements{H.N. acknowledges funding from Research Corp. and the NSF
(AST-0307571).  Funding for the creation and distribution of the SDSS Archive ha
s been provided by the Alfred P. Sloan Foundation, the Participating
Institutions, NASA, the NSF, DOE, the Japanese Monbukagakusho, and the 
Max Planck Society.  The SDSS Web site is http://www.sdss.org/.}


\end{document}